# Unsupervised Learning for Neural Network-based Polar Decoder via Syndrome Loss

Chieh-Fang Teng, *Student Member, IEEE*, and An-Yeu (Andy) Wu, *Fellow, IEEE*

*Abstract*—With the rapid growth of deep learning in many fields, machine learning-assisted communication systems had attracted lots of researches with many eye-catching initial results. At the present stage, most of the methods still have great demand of massive labeled data for supervised learning. However, obtaining labeled data in the practical applications is not feasible, which may result in severe performance degradation due to channel variations. To overcome such a constraint, syndrome loss has been proposed to penalize non-valid decoded codewords and achieve unsupervised learning for neural network-based decoder. However, it cannot be applied to polar decoder directly. In this work, by exploiting the nature of polar codes, we propose a modified syndrome loss. From simulation results, the proposed method demonstrates that domain-specific knowledge and know-how in code structure can enable unsupervised learning for neural network-based polar decoder.

## I. Introduction

WITH more and more revolutionized breakthroughs in the field of computer vision and natural language processing, machine learning-assisted communication systems have also attracted a lot of researchers in this newly emerging field. Most of the well-designed networks are either as the replacements for certain blocks [1]-[6] or as an end-to-end solutions [7]-[8]. In [1]-[2], convolutional neural networks are exploited for a powerful modulation classification, outperforming conventional approaches. For channel decoding, neural network-based decoders for high-density parity-check (HDPC) codes and polar codes are proposed in [3]-[5], which assign trainable weights to the edges and improve the convergence speed with better performance. In [6], a neural network-based equalizer is proposed, which not only eliminates channel fading, but also exploits the code structure with utilization of coding gain in advance. For end-to-end optimization, by replacing the whole system with neural networks, the authors in [7]-[8] try to break the conventional rules of independent block design by jointly optimizing communication systems.

Though neural networks can achieve promising performance in the simulation experiments, most of the methods are based on supervised learning as shown in Fig. 1(a), which neglects the feasibility of obtaining labeled training data in the practical applications. Unfortunately, without the precious training data for accurate channel estimation, the performance degrades severely under time-varying channels [9]-[10]. Therefore, unsupervised learning plays an important role in overcoming the challenge of online channel adaptation.

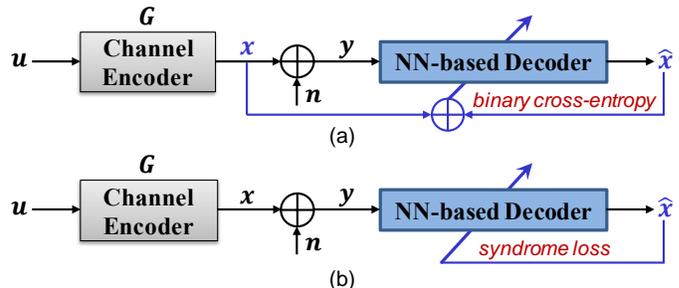

Fig. 1. Training process for neural network-based decoder: (a) Supervised learning with binary cross-entropy loss. (b) Unsupervised learning with syndrome loss.

Recently, syndrome loss was proposed in [11]. It penalizes the decoder for producing non-valid codewords, and it can be used to train neural network-based decoder for Bose-Chaudhuri-Hocquenghem (BCH) codes and low-density parity-check (LDPC) without prior knowledge of the transmitted codewords as shown in Fig. 1(b). Therefore, the decoder can be trained by unsupervised learning, making it a promising tool for online channel adaptation. However, this approach demands the decoder to output the soft estimation of the codewords. Thus, it can be examined by parity-check matrix to produce syndrome loss. Unfortunately, polar decoder [12], whose outputs are source bits without the definition of parity-check matrix, cannot be directly applied syndrome loss for unsupervised learning.

In this work, by exploiting the nature of polar codes, we propose a modified syndrome loss, frozen-bit syndrome loss, which enables unsupervised learning to be used for neural network-based polar decoder without the requirements of labeled data. Simulation results are carried out to demonstrate its effectiveness.

The remaining part of this letter is organized as follows. Section II briefly reviews the syndrome loss and neural network-based polar decoder. Section III derives the proposed frozen-bit syndrome loss with detailed training process. The numerical experiments and analyses are shown in Section IV. Finally, Section V concludes this letter.

## II. Preliminaries

### A. Syndrome Loss

In [11], the authors introduced the syndrome loss, which penalizes the decoder for producing outputs that do not correspond to valid codewords. In communication systems, the transmitter encodes a $K$-bit message $\boldsymbol{u} \in \text{GF}(2)^K$ by using a generator matrix $\boldsymbol{G} \in \text{GF}(2)^{N \times K}$ to obtain an $N$-bit codeword $\boldsymbol{c} = \boldsymbol{G}\boldsymbol{u} \in \text{GF}(2)^N$. After transforming to a bipolar format $\boldsymbol{x} = 1 - 2\boldsymbol{c} \in \{-1, 1\}^N$, the codeword is transmitted over the channel. Then, the decoder will estimate $\boldsymbol{x}$ from the received noisy signal $\boldsymbol{y} = \boldsymbol{x} + \boldsymbol{n}$, where $\boldsymbol{n}$ is the additive white Gaussian noise (AWGN) with variance $\sigma^2$. The estimated bipolar codeword, $\hat{\boldsymbol{x}} = \text{sign}(\boldsymbol{s})$, is found by taking the hard decision of soft output from decoder $\boldsymbol{s} \in \mathbb{R}^N$, and the corresponding estimated binary codeword is $\hat{\boldsymbol{c}} = 0.5 - 0.5\hat{\boldsymbol{x}}$.

For a linear code, the estimated binary codeword $\hat{\boldsymbol{c}}$ can be examined by a parity-check matrix $\boldsymbol{H} \in \text{GF}(2)^{(N-K) \times N}$, and the syndrome is defined as the product $\boldsymbol{H}\hat{\boldsymbol{c}} \in \text{GF}(2)^{N-K}$. For a valid codeword, the syndrome must contain only 0. Therefore, the syndrome can be used to check if the decoder has successfully produced a valid codeword. Based on this concept, a differentiable soft syndrome can be defined as follows:

$$\text{softsynd}(\boldsymbol{s})_i = \min_{j \in \mathcal{M}(i)} |s_j| \prod_{j \in \mathcal{M}(i)} \text{sign}(s_j), \quad (1)$$

where $\mathcal{M}(i)$ is the set of entries in the $i$th row of $\boldsymbol{H}$ equal to 1 and this equation is extended from the check node update equation in min-sum decoding algorithm.

To maximize each entry in the soft syndrome, the syndrome loss can be constructed to penalize the negative ones and given by:

$$\mathcal{L}_{\text{synd}}(\boldsymbol{s}) = \frac{1}{N-K} \sum_{i=0}^{N-K-1} \max(1 - \text{softsynd}(\boldsymbol{s})_i, 0). \quad (2)$$

Therefore, the loss function can be calculated without the knowledge of transmitted codeword $\boldsymbol{c}$ and backpropagated for the training of neural network-based decoder under unsupervised learning. On the other hand, the commonly used loss function for supervised binary classification is binary cross-entropy, which requires the codeword $\boldsymbol{c}$ for calculation as follows:

$$\mathcal{L}_{\text{BCE}}(\boldsymbol{c}, \boldsymbol{s}) = \frac{1}{N} \sum_{i=0}^{N-1} [c_i \log g(-s_i) + (1-c_i) \log(1 - g(-s_i))], \quad (3)$$

where $g$ is the sigmoid function.

*B. Polar Codes with Neural Network-Based Polar Decoder*

To construct an $(N, K)$ polar code, the $N$-bit message $\boldsymbol{u}$ is recursively constructed from a polarizing matrix $\boldsymbol{F} = \begin{bmatrix} 1 & 0 \\ 1 & 1 \end{bmatrix}$ by $\log_2 N$ times to exploit channel polarization [12]. As $N \to \infty$, the synthesized channels tend to two extremes: the noisy channels and noiseless channels. Therefore, the $K$ information bits are assigned to the $K$ most reliable bits in $\boldsymbol{u}$ and the remaining $(N - K)$ bits are referred to as frozen bits with the assignment of zeros. Then, the $N$-bit transmitted codeword $\boldsymbol{c}$ can be generated by multiplying $\boldsymbol{u}$ with generator matrix $\boldsymbol{G}$ as follows:

$$\boldsymbol{c} = \boldsymbol{G}\boldsymbol{u} = \boldsymbol{F}^{\otimes n} \boldsymbol{B} \boldsymbol{u}, n = \log_2 N. \quad (4)$$

$\boldsymbol{F}^{\otimes n}$ is the $n$-th Kronecker power of $\boldsymbol{F}$, and $\boldsymbol{B}$ represents the bit-reversal permutation matrix.

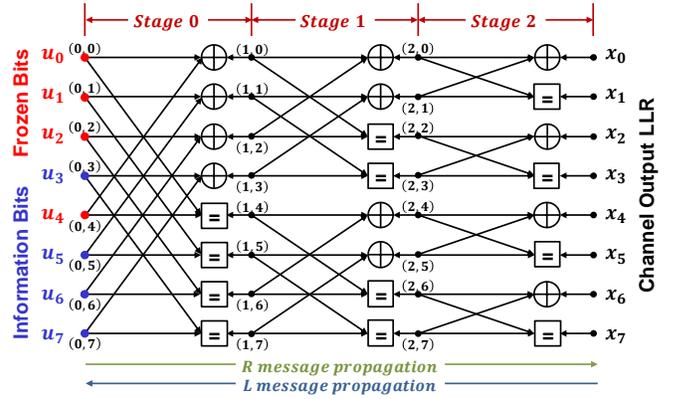

Fig. 2. Factor graph of polar codes with $N = 8. A = \{3, 5, 6, 7\}$ and $A^c = \{0, 1, 2, 4\}$.

Belief propagation (BP) is a widely used algorithm for polar decoder [4]-[5]. There are two types of log likelihood ratios (LLRs) iteratively updated on the factor graph, namely left-to-right message $R_{i,j}^{(t)}$ and right-to-left message $L_{i,j}^{(t)}$, where node $(i, j)$ represents $j$-th node at the $i$-th stage and $t$ indicates the $t$-th iteration as shown in Fig. 2. In [4]-[5], by taking advantage of DL, a neural network-based belief propagation (NN-BP) is proposed with assigned weight on the factor graph. After training, the weights function as the scaling factor for the importance of messages and the iterative decoding procedure is revised as:

$$\begin{cases} L_{i,j}^{(t)} = \alpha_{i,j}^{(t)} \cdot g\left(L_{i+1,j}^{(t-1)}, L_{i+1,j+N/2^i}^{(t-1)} + R_{i,j+N/2^i}^{(t)}\right), \\ L_{i,j+N/2^i}^{(t)} = \alpha_{i,j+N/2^i}^{(t)} \cdot g\left(R_{i,j}^{(t)}, L_{i+1,j}^{(t-1)}\right) + L_{i+1,j+N/2^i}^{(t-1)}, \\ R_{i+1,j}^{(t)} = \beta_{i+1,j}^{(t)} \cdot g\left(R_{i,j}^{(t)}, L_{i+1,j+N/2^i}^{(t-1)} + R_{i,j+N/2^i}^{(t)}\right), \\ R_{i+1,j+N/2^i}^{(t)} = \beta_{i+1,j+N/2^i}^{(t)} \cdot g\left(R_{i,j}^{(t)}, L_{i+1,j}^{(t-1)}\right) + R_{i,j+N/2^i}^{(t)}, \end{cases} \quad (5)$$

where $\alpha_{i,j}^{(t)}$ and $\beta_{i,j}^{(t)}$ denote the right-to-left and left-to-right trainable scaling parameters, respectively. Finally, after $T$ iterations, the estimation of $\hat{\boldsymbol{u}}$ is decided by:

$$\hat{u}_j = \begin{cases} 0, & \text{if } L_{1,j}^{(T)} \geq 0, \\ 1, & \text{if } L_{1,j}^{(T)} < 0. \end{cases} \quad (6)$$

For more details, please refer to [4]-[5].

### III. PROPOSED SYNDROME LOSS FOR POLAR DECODER

For a BCH code or a LDPC code, belief propagation decoding is iteratively performed on the bipartite graph, which is constructed from the well-defined parity-check matrix. Besides, the output of decoder is a soft estimation of codeword and can be directly checked by the matrix, which meets with the requirements for using syndrome loss.

However, polar decoder directly estimates the message $\boldsymbol{u}$ from (6) without providing the definition of $\boldsymbol{H}$, which restricts the usage of syndrome loss. To address this issue, by exploiting the nature of polar code, we derive a modified syndrome loss with suitable parity-check matrix for unsupervised learning.

*A. Frozen-Bit Syndrome Loss*

Consequently, for polar codes, we need to define the parity-check matrix $\boldsymbol{H}$ with suitable output $\boldsymbol{s}$ from decoder to produce soft syndrome in (1) for penalization. Though polar code does not provide a parity-check matrix inherently, it has a special characteristic, the frozen bits, which are set to 0 and allow us to derive the parity-check matrix.

Firstly, the codeword $c = Gu$ can be inverted as:

$$u = G^{-1}c = (F^{\otimes n}B)^{-1}c = (F^{-1})^{\otimes n}B^{-1}c, \quad (7)$$

where $F^{-1} = F$ and $B^{-1} = B$. Thus, (7) can be simplified as:

$$u = (F^{-1})^{\otimes n}B^{-1}c = F^{\otimes n}Bc = Gc. \quad (8)$$

Then, based on the characteristic of polar codes that the frozen bits are set to 0, $u$ with indices of the frozen bits must be 0. Thus, (8) can be further derived as:

$$u_{A^c} = G_{A^c}c = 0, \quad (9)$$

where $A \subseteq \{0, 1, ..., N-1\}$ denotes the set of indices for the information bits and $A^c$ is its complement to represent the set of indices for the frozen bits. Consequently, we can conclude that the parity-check matrix $H$ for polar code is as follows:

$$H = G_{A^c}, \quad (10)$$

which is formed from the rows of $G$ with indices in $A^c$. The soft output from decoder $s$ can be obtained as:

$$s_j = L_{n+1,j}^{(T)} + R_{n+1,j}^{(T)}, \forall j \in \{0, ..., N-1\}, \quad (11)$$

which is used for the calculation of syndrome loss.

### B. Training Process Overview

Now, based on the proposed syndrome loss, the unsupervised learning for neural network-based polar decoder is provided in Algorithm 1. Firstly, the random generated message $u$ can be encoded by (4) and then the received signal $y$ is obtained by transmitting the codeword over the channel with added noise. The log-likelihood ratios (LLRs), as the input for the BP decoder, can be transformed from $y$ and given by:

$$llr = 2y/\sigma^2. \quad (12)$$

For each iteration of training, a mini-batch of size $M$ of LLRs are operated in parallel to improve the convergence speed and ensure the stability of training process. After $T$ iterations of BP decoding, the soft output from decoder can be obtained from updated messages $L$ and $R$ according to (11). Then, the trainable scaling parameters $\alpha$ and $\beta$ are optimized through gradient descent on the proposed syndrome loss. In this work, the adopted algorithm for optimization is stochastic gradient descent (SGD), which iteratively updates the trainable parameters based on the gradient of the loss function as follows:

$$\theta^{(j+1)} = \theta^{(j)} - \eta \nabla_\theta \mathcal{L}_{\text{synd}}(\theta^{(j)}), \theta = \{\alpha, \beta\}, \quad (13)$$

where $\theta$ is the set of trainable parameters, $\eta > 0$ is the learning rate, and $\nabla_\theta \mathcal{L}_{\text{synd}}$ is the gradient of the proposed syndrome loss function. After reaching a fixed number of training iterations or meeting some stop criterion, the well-trained parameters can be obtained, which achieves unsupervised learning without the knowledge of labeled data.

## IV. SIMULATION RESULTS

In the simulation experiments, we evaluate the proposed frozen-bit syndrome loss on RNN-based polar decoder [5]. The

**Algorithm 1:** Unsupervised Learning for Neural Network-based Polar Decoder

**Input:** $llr, A, T, \eta$
**Output:** $\alpha, \beta$
1: $\{\alpha, \beta\} \leftarrow 1$
2: **while** training stop criterion not met **do**
3:    $L, R \leftarrow$ initialize the NN-BP decoder($llr, A$)
4:    $L, R \leftarrow$ NN-BP decoder($L, R, \alpha, \beta, T$)
5:    **for** $j = 0$ to $N - 1$ **do**
6:       $s_j = L_{n+1,j}^{(T)} + R_{n+1,j}^{(T)}$
7:    $\{\alpha, \beta\} \leftarrow \text{SGD}(\alpha, \beta, \mathcal{L}_{\text{synd}}(s), \eta)$

TABLE I
SIMULATION PARAMETERS

| Encoding | Polar code (64,32) |
|---|---|
| Iteration | 5 |
| Modulation category | BPSK |
| Signal to Noise Ratio (SNR) | 0, 1, 2, 3, 4, 5 (dB) |
| Training codeword/SNR | 60,000 |
| Testing codeword/SNR | 151,200 |
| Mini-batch size ($M$) | 3,600 |
| Learning rate | 0.03 |
| Validation ratio | 0.2 |
| Optimizer | Stochastic gradient descent (SGD) |
| Training and testing environment | Deep learning library of TensorFlow with NVIDIA GTX 1080 Ti GPU |

adopted recurrent architecture forces the network to learn shared weights among different iterations, which is helpful for reducing memory overhead and hardware complexity. The simulation setup is summarized in Table I.

### A. Convergence Speed for RNN-based Polar Decoder with Frozen-bit Syndrome Loss

In this experiment, we evaluate stability and convergence speed by comparing the evolutions of the validated losses and validated frame error rate (FER) during the first 50 training epochs. Firstly, we compare the loss between supervised learning and unsupervised learning, which use binary cross-entropy and the proposed syndrome loss, respectively.

In Fig. 4, the losses are averaged over 30 seeds and the shaded areas around the curves correspond to one standard deviation in each direction. In general, we may consider that supervised learning has better convergence speed over unsupervised learning due to the demanded ground truth of message $u$ for training. However, to our surprise, both methods almost have the same convergence speed with small variance, which demonstrates the stability of the proposed syndrome loss for unsupervised learning.

Then, we compare the more important measurement metric of FER to evaluate the improvement of FER contributed from the decline of these two different losses. In Fig. 5, we can find out that the decline of binary cross-entropy has more direct impact on the improvement of FER. This is reasonable because the demanded message $u$ for supervised learning can accurately correct the output value of each bit. On the other hand, some wrongly decoded frames may not be penalized by the syndrome loss.

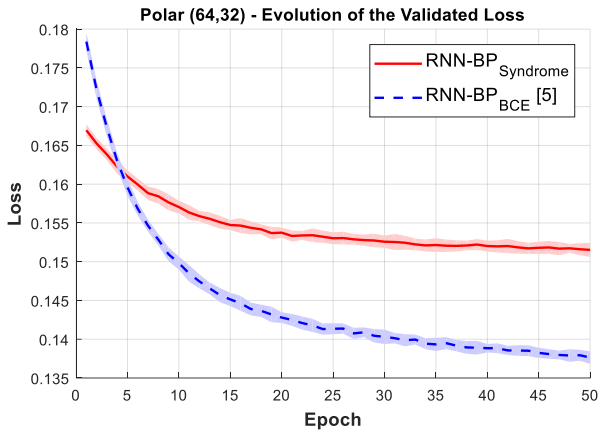

Fig. 4. Evolution of the validated loss between the proposed frozen-bit syndrome loss and binary cross-entropy (BCE).

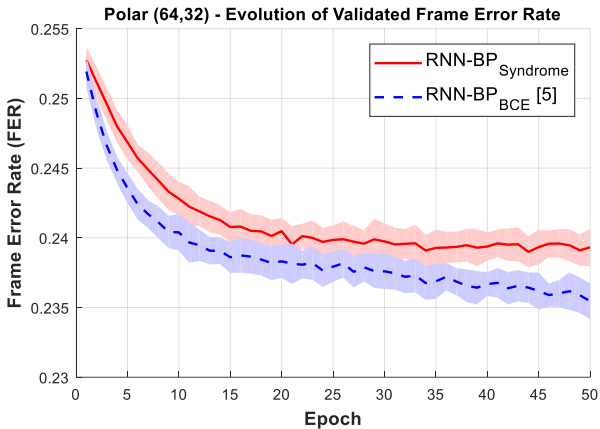

Fig. 5. Evolution of the validated frame error rate between the proposed frozen-bit syndrome loss and binary cross-entropy (BCE).

### B. Performance for RNN-based Polar Decoder with Frozen-bit Syndrome Loss

In this experiment, we further compare FER between binary cross-entropy and the proposed syndrome loss under different SNR. Besides, the performance of conventional BP is also measured for comparison.

From Fig. 6, we can see that supervised learning achieve the best performance because it has the ground truth of message $u$ for training, which has been adopted and evaluated in prior works [3]-[5]. On the other hand, the performance of unsupervised learning with proposed frozen-bit syndrome loss only has slight improvement over conventional BP and is worse than supervised learning. This result may come from another reason that only the part of the frozen bits is constrained, which is not comprehensive enough for training the whole polar factor graph. However, it still proves that the proposed syndrome loss can exploit the nature of polar codes and achieve unsupervised learning when the labeled data is not provided.

## V. CONCLUSION

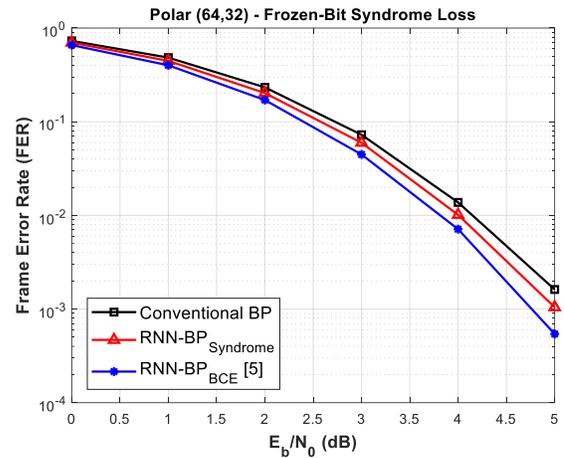

Fig. 6. Comparison of FER performance between the proposed frozen-bit syndrome loss and binary cross-entropy (BCE).

In this paper, by exploiting the nature of polar codes, we propose a modified syndrome loss, which makes unsupervised learning possible for neural network-based polar decoder. We believe that the domain-specific syndrome loss will have many useful applications and is a powerful tool for the paradigm of machine learning assisted communication systems to overcome many realistic problems, such as channel variations.